\documentclass[showpacs,preprintnumbers,amsmath,amssymb,twocolumn]{revtex4}

\usepackage{amsmath}
\usepackage{cases}
\usepackage{amsfonts}

\usepackage{graphicx}
\usepackage{dcolumn}
\usepackage{bm}
\usepackage{amssymb}
\usepackage{array}
\usepackage{longtable}
\usepackage{hhline}

\begin{document}

\preprint{APS/123-QED}

\title{Random walks on weighted networks}

\author{Zhongzhi Zhang$^{1,2}$}
\email{zhangzz@fudan.edu.cn}
\homepage{http://homepage.fudan.edu.cn/~zhangzz/}

\author{Tong Shan$^{1,2}$}

\author{Guanrong Chen$^{3}$}

\affiliation {$^{1}$ School of Computer Science, Fudan University, Shanghai
200433, China}
\affiliation {$^{2}$ Shanghai Key Lab of Intelligent Information
Processing, Fudan University, Shanghai 200433, China}
\affiliation {$^{3}$ Department of Electronic Engineering, City University of Hong Kong, Hong Kong SAR, China}




\date{\today}

\begin{abstract}
Random walks constitute a fundamental mechanism for a large set of dynamics taking place on networks. In this article, we study random walks on weighted networks with an arbitrary degree distribution, where the weight of an edge between two nodes has a tunable parameter. By using the spectral graph theory, we derive analytical expressions for the stationary distribution, mean first-passage time (MFPT), average trapping time (ATT), and lower bound of the ATT, which is defined as the average MFPT to a given node over every starting point chosen from the stationary distribution. All these results depend on the weight parameter, indicating a significant role of network weights on random walks. For the case of uncorrelated networks, we provide explicit formulas for the stationary distribution as well as ATT. Particularly, for uncorrelated scale-free networks, when the target is placed on a node with the highest degree, we show that ATT can display various scalings of network size, depending also on the same parameter. Our findings could pave a way to delicately controlling random-walk dynamics on complex networks.

\end{abstract}


\maketitle


\section{Introduction}

Random walks on complex networks have attracted increasing interest in the past few years~\cite{MeKl04,BuCa05} because of their wide applications in a large variety of scientific fields~\cite{We1994}. Thus far, random walks on binary networks have been extensively studied, leading to important understanding of this paradigmatic dynamical process and uncovering the effects of the underlying network structure on the behavior of random-walk dynamics~\cite{CoBeTeVoKl07}. For example, general solutions for mean first-passage time (MFPT) from one node to another in a general network have been found by using different methods~\cite{NoRi04,Vo11,ZhAlHoZhCh11}. The MFPT from node $i$ to node $j$, denoted by $F_{ij}$, is defined as the expected time for a walker starting from node $i$ to first reach node $j$. This quantity is required in various considerations~\cite{Re01}. Recently, average trapping time (ATT), $\langle F_j \rangle$, for the trapping problem~\cite{Mo69}, a particular random walk defined as the average of $F_{ij}$ over all starting points $i$ to a given trap node $j$, has received considerable attention~\cite{KaBa02,Ag08,HaRo08,KiCaHaAr08,ZhQiZhXiGu09,ZhXiZhLiGu09,ZhGuXiQiZh09,AgBu09,TeBeVo09,ZhLiGaZhGuLi09,BeTuKo10,LiWuZh10,AgBuMa10,ZhWuCh11,ZhYaLi12,MeAgBeVo12,WuLiZhCh12,LiJuZh12,HwLeKa12}, for this quantity can be applied to characterize transport efficiency. There are different approaches for evaluating $\langle F_j \rangle$, showing nontrivial (dependence) scalings of MFPT $\langle F_j \rangle$ with network sizes, which strongly depend on network structural properties such as power-law degree distribution~\cite{KiCaHaAr08,ZhQiZhXiGu09,AgBu09,TeBeVo09}, fractality~\cite{ZhGuXiQiZh09}, and modularity~\cite{ZhLiGaZhGuLi09,ZhYaLi12}.

It is well known that in addition to the above-mentioned important structural properties, most real-life networks exhibit a large heterogeneity in the distribution of edge weights, characterizing the intensive interactions between node pairs~\cite{BaBaPaVe04,MaAlBa05}. The inhomogeneous weights have a substantial effect on various dynamical processes running on a network, including zero range processes~\cite{TaLiZh06}, synchronization~\cite{ZhMoKu06,Ko07}, cascading failures~\cite{WaCh08}, mass-aggregation~\cite{KwYoKi08}, diffusive processes~\cite{BaPa10}, bimolecular chemical reactions~\cite{KwChKi10}, diffusion-annihilation processes~\cite{ZhZhGuZh11}, voter behavior~\cite{BaCaSa11}, and traffic fluctuation~\cite{ZhZhZhGuZhCh12}. Thus far, a comprehensive analysis about random walks of weighted networks is still lacking, although it is suggested that weight heterogeneity could play an essential role in random-walk dynamics.

In this paper, we present a rather comprehensive study of random walks on a generic weighted network, where the weight $w_{ij}$ of the edge linking nodes $i$ and $j$ has strong correlations with the degrees, $d_{i}$ and $d_{j}$, of nodes $i$ and $j$, and is assumed to be $w_{ij}=(d_{i}d_{j})^\theta$ with $\theta$ being a controllable parameter. Such a form of edge weights can be observed in various real weighted networks~\cite{BaBaPaVe04,MaAlBa05}. Based on spectral graph theory~\cite{Ch97}, we will derive analytically the formulas of the stationary distribution, MFPT $F_{ij}$ from node $i$ to node $j$, ATT $\langle F_j \rangle$ to a given target node $j$ for a walker starting from a source node selected from the stationary distribution, as well as a lower bound for $\langle F_j \rangle$. The obtained results show that all these quantities are strongly affected by network weights characterized by parameter $\theta$, which indicates the significant impact of weights on random walks. For the special case of uncorrelated networks~\cite{DoGoMe08}, we will provide succinct expressions for some interesting quantities. Particularly, for the trapping problem on an uncorrelated scale-free network, when the perfect trap is placed at a hub node with the largest degree, by tuning the weight parameter $\theta$, ATT can exhibit different scalings with network sizes. Our work thus might have practical implications for controlling random walks on real-world networks.

\section{Formulation of the problem} \label{RanWalk}

Most existing works on random walks are confined to binary networks~\cite{NoRi04}. For a connected binary network $G$ with $N$ nodes and $M$ edges, where  the nodes are labeled by $1,2,3,\ldots, N$, its connectivity is represented by the adjacency matrix $A$, in which the entry $a_{ij}=1$ (or 0) if nodes $i$ and $j$ are (not) adjacent. The degree of node $i$ is $d_i=\sum_{j=1}^{N} a_{ij}$, and the diagonal degree matrix $D$ of network $G$ is defined as follows: the $i$th diagonal entry is $d_i$, while all the nondiagonal entries are equal to zero. It is easy to verify that the total degree of all $N$ nodes is $K=2M= \sum_{j=1}^{N}d_i$, and the average node degree is $\langle d \rangle=2M /N$.

For discrete-time random walks defined on a binary network, at each time step the particle (walker) starting from its current position moves to each of its neighboring nodes with the same probability. Such a stochastic process is characterized by the transition matrix $P=D^{-1}A$~\cite{Vo11}, in which the entry $p_{ij}= a_{ij}/d_i$ presents the jumping probability from node $i$ to node $j$ in one step. This process can be described by an ergodic Markov chain~\cite{KeSn76,AlFi99}, whose stationary distribution is $\pi=(\pi_1, \pi_2,\ldots, \pi_N)^\top$, where $\pi_i=d_i/K$. It is clear that $\pi^{\top}P=\pi^{\top}$ and $\sum_{i=1}^{N}\pi_i=1$.

The above-defined random walks can be easily extended to be on weighted networks~\cite{WuXuWuWa07}. Mathematically, topological and weighted properties of a weighted network are described by a generalized adjacency matrix $W$, whose entry $w_{ij}$ specifies the weight of the edge connecting  nodes $i$ and $j$. In this paper, we focus on an undirected network having symmetric nonnegative weights $w_{ij}=w_{ji}\geq 0$. Moreover, we assume the weight of the edge between nodes $i$ and $j$ to be $w_{ij}=(d_{i}d_{j})^\theta$, where $\theta$ is a controllable parameter.  Our assumption is based on empirical work on realistic networks, including scientific collaboration networks~\cite{BaBaPaVe04}, metabolic networks~\cite{MaAlBa05}, and airport networks~\cite{BaBaPaVe04,MaAlBa05}. Moreover, the strength $s_{i}$ of node $i$ is defined by $s_{i}=\sum_{j\in \Omega_i} w_{ij}$~\cite{BaBaVe04}, where the sum runs over the set $\Omega_i$ of the neighboring nodes of $i$.

Notice that random walks occurring on weighted networks are biased. In the process of a random walk, the transition probability $p_{ij}$ from node $i$ to $j$ is given by $p_{ij}=w_{ij}/s_{i}$, which constitutes an entry of the transition matrix $P=S^{-1}W$ for the biased random walk, where $S$ is the diagonal strength matrix with its $i$th diagonal entry equal to the strength $s_{i}$ of node $i$. In the following, let $s$ denote the sum of strengths for all the $N$ nodes, i.e., $s=\sum_{i=1}^{N}s_{i}=\sum_{i=1}^{N}\sum_{j=1}^{N}w_{ij}$. It is clear that the transition matrix $P$ is a stochastic matrix.

Next, we will show that many quantities related to the biased random walks are encoded in the transition matrix $P$ and thus can be derived from $P$. Generally, except regular networks, $P$ is asymmetric. So, we introduce the matrix
\begin{equation}\label{Trans01}
\Gamma=S^{-\frac{1}{2}}WS^{-\frac{1}{2}}=S^{\frac{1}{2}}PS^{-\frac{1}{2}},
\end{equation}
which is real and similar to $P$ and thus has the same set of eigenvalues as $P$. Furthermore, if $\psi$ is an eigenvector of matrix $\Gamma$ associated with eigenvalue $\lambda$, then $S^{-\frac{1}{2}}\psi$ is an eigenvector of $P$ corresponding to eigenvalue $\lambda$.

It is easy to verify that for any given node $i$, $\sum_{j=1}^n p_{ij}=1$ always holds. Thus, 1 is an eigenvalue of transition matrix $P$, as well as matrix $\Gamma$. Moreover, it is the greatest eigenvalue with single degeneracy.
For eigenvalue 1 with a corresponding normalized eigenvector $w$, one has $\Gamma w=w$, which together with Eq.~(\ref{Trans01}) gives $P(S^{-\frac{1}{2}}w)=(S^{-\frac{1}{2}}w)$. Because $P1=1$, one has $S^{-\frac{1}{2}}w=1$. Thus, $w$ can be determined explicitly as
\begin{equation}\label{Trans00}
w=(w_{1},w_{2},\ldots,w_{N})^\top =\left(\sqrt{\frac{s_1}{s}},\sqrt{\frac{s_2}{s}},\ldots,\sqrt{\frac{s_N}{s}}\right)^\top.
\end{equation}

Equation~(\ref{Trans01}) shows that matrix $\Gamma$ is real and symmetric; thus all its eigenvalues are real. Let $\lambda_1$, $\lambda_2$, $\lambda_3$, $\cdots$, $\lambda_N$ be the $N$ eigenvalues of matrix $\Gamma$ for a network of size $N$, rearranged as
$1=\lambda_1>\lambda_2 \geq \lambda_3 \geq \ldots \geq \lambda_N \geq -1$, and let $\psi_1$, $\psi_2$, $\psi_3$, $\ldots$, $\psi_N$ denote the corresponding normalized, real-valued, and mutually orthogonal eigenvectors. Let $\psi_i=(\psi_{i1},\psi_{i2},\ldots,\psi_{iN})^{\top}$ and $\psi_i$ be the $i$th column vector of matrix $\Psi $. Evidently, $\Psi$ is an orthogonal matrix, satisfying
\begin{equation}\label{Trans02}
\Psi \Psi^{\top}=\Psi^{\top}\Psi=I;
\end{equation}
that is,
\begin{equation}\label{Trans03}
\sum_{k=1}^{N}\psi_{ik}\psi_{jk}=\sum_{k=1}^{N}\psi_{k i}\psi_{kj}=\begin{cases}
1, &{\rm if} \quad i=j,\\
0, &{\rm otherwise}.
\end{cases}
\end{equation}

According to the properties of real symmetric matrices, one has
\begin{equation}\label{Trans04}
\Psi^{\top} \Gamma \Psi={\rm diag}[\lambda_1, \lambda_2, \ldots, \lambda_N],
\end{equation}
which, together with Eq.~(\ref{Trans02}), leads to
\begin{equation}\label{Trans05}
\Gamma=\Psi{\rm diag}\,[\lambda_1, \lambda_2, \ldots, \lambda_N]\,\Psi^{\top}\,.
\end{equation}
Equation~(\ref{Trans05}) means that the entry $\tau_{ij}$ of matrix $\Gamma$ has the following spectral form:
\begin{equation}\label{Trans06}
\tau_{ij}=\sum_{k=1}^{N}\lambda_k \psi_{k i} \psi_{kj}.
\end{equation}

We now express matrix $P$ in terms of $\Psi$ and $S$. From Eqs.~(\ref{Trans01}) and~(\ref{Trans02}) one  obtains
\begin{equation}\label{Trans07}
P=S^{-\frac{1}{2}}\Gamma S^{\frac{1}{2}}=S^{-\frac{1}{2}}\Psi{\rm diag}\,[\lambda_1, \lambda_2, \ldots, \lambda_N]\,\Psi^{\top}S^{\frac{1}{2}}.
\end{equation}
Let $P^t$ be the $t$th power of matrix $P$, whose $ij$th entry denoted by $(p^{t})_{ij}$ represents the probability for a walker to start from node $i$ and reach node $j$ in $t$ steps.
From Eq.~(\ref{Trans07}), one has
\begin{equation}\label{Trans08}
P^t = S^{-\frac{1}{2}}\left(\sum_{k=1}^{N}\lambda^t_k \psi_k \psi_k^{\top}\right) S^{\frac{1}{2}}
\end{equation}
and
\begin{equation}\label{Trans09}
(p^{t})_{ij}=\sum_{k=1}^{N}\lambda^t_k\psi_{ki}\psi_{kj}\sqrt{\frac{s_{j}}{s_{i}}}\,.
\end{equation}
In the limit, one has
\begin{equation}\label{Trans10}
\lim_{t \rightarrow \infty}(p^{t})_{ij}=\psi_{1i}\psi_{1j}\sqrt{\frac{s_{j}}{s_{i}}}=\frac{s_{j}}{s}\,,
\end{equation}
where Eq.~(\ref{Trans00}) and the relation $w=\psi_{1}$ were used.
Therefore, the stationary distribution for a random walk on weighted networks is
\begin{equation}\label{Trans11}
\pi=(\pi_1, \pi_2,\ldots, \pi_N)^\top=\left(\frac{s_{1}}{s}, \frac{s_{2}}{s},\ldots, \frac{s_{N}}{s}\right)^\top,
\end{equation}
which obviously depends on the strengths of the nodes and thus the weights of the edges.

\section{MFPT from one node to another \label{model}}

MFPT is one of the most important quantities related to random walks, since it contains   useful information about the random-walk dynamics~\cite{Re01}. In what follows, we will derive MFPT for a random walk from one node to another in a weighted network, by using the formalism of generating functions~\cite{Wi94}.

Let $(q^t)_{ij}$ represent the probability for a walker starting from node $i$ to reach node $j$ for the first time precisely in $t$ steps. This first passage  probability $(q^t)_{ij}$ relates to the transition probability $(p^t)_{ij}$ by
\begin{equation}\label{MFPT01}
(p^{t})_{ij}=\sum_{s=0}^{t}(q^{s})_{ij}(p^{t-s})_{jj}\,.
\end{equation}

Let $F_{ij}$ denote the MFPT for a random walk on a weighted network from node $i$ to $j$.
Then,
\begin{equation}\label{MFPT02}
F_{ij}=\sum_{t=1}^{\infty}t(q^{t})_{ij}.
\end{equation}
To find $F_{ij}$, we introduce the following generating functions:
\begin{equation}\label{MFPT03}
\tilde{Q}_{ij}(x)=\sum_{t=0}^{\infty}(q^{t})_{ij} x^{t}\,
\end{equation}
and
\begin{equation}\label{MFPT04}
\tilde{P}_{ij}(x)=\sum_{t=0}^{\infty}(p^{t})_{ij} x^{t}\,,
\end{equation}
where $|x|<1$.
The above two equations together yield
\begin{equation}\label{MFPT05}
\tilde{Q}_{ij}(x)=\frac{\tilde{P}_{ij}(x)}{\tilde{P}_{jj}(x)}\,.
\end{equation}
By definition, the MFPT $F_{ij}$ can be evaluated as 
\begin{equation}\label{MFPT06}
F_{ij}=\sum_{t=1}^{\infty}t(q^{t})_{ij}=\frac{\rm d}{{\rm d} x}\tilde{Q}_{ij}(x) \bigg |_{x=1}.
\end{equation}
Thus, our goal is reduced to evaluating $\tilde{Q}_{ij}(x)$ and then differentiating it.

Plugging Eq.~(\ref{Trans09}) into Eq.~(\ref{MFPT04}) yields
\begin{eqnarray}\label{MFPT07}
\tilde{P}_{ij}(x)&=&\sum_{t=0}^{\infty}\sum_{k=1}^{N}(\lambda_{k}x)^{t}\psi_{ki}\psi_{kj}\sqrt{\frac{s_{j}}{s_{i}}}\notag\\
&=&\sum_{k=1}^{N}\psi_{ki}\psi_{kj}\sqrt{\frac{s_{j}}{s_{i}}}+\sum_{k=1}^{N}\sum_{t=1}^{\infty}(\lambda_{k}x)^{t}\psi_{ki}\psi_{kj}\sqrt{\frac{s_{j}}{s_{i}}}\notag\\
&=&\psi_{1i}\psi_{1j}\sqrt{\frac{s_{j}}{s_{i}}}\sum_{t=1}^{\infty}x^{t}+\sum_{k=2}^{N}\left[\sum_{t=1}^{\infty}(\lambda_{k}x)^{t}\right]\psi_{ki}\psi_{kj}\sqrt{\frac{s_{j}}{s_{i}}}\notag\\
&=&\pi_{j}\frac{1}{1-x}+\sum_{k=2}^{N}\frac{1}{1-\lambda_{k}x}\psi_{ki}\psi_{kj}\sqrt{\frac{s_{j}}{s_{i}}},
\end{eqnarray}
where we have used Eqs.~(\ref{Trans03}) and~(\ref{Trans11}).
In a similar way, $\tilde{P}_{jj}(x)$ can be derived as follows:
\begin{eqnarray}\label{MFPT08}
\tilde{P}_{jj}(x)&=&\sum_{t=0}^{\infty}\sum_{k=1}^{N}(\lambda_{k}x)^{t}\psi_{kj}^{2}\notag\\
&=&\sum_{k=1}^{N}(\lambda_{1}x)^{0}\psi_{kj}^{2}+\sum_{k=1}^{N}\sum_{t=1}^{\infty}(\lambda_{k}x)^{t}\psi_{kj}^{2}\notag\\
&=&\sum_{k=1}^{N}\psi_{kj}^{2}+\psi_{1j}^{2}\sum_{t=1}^{\infty}(\lambda_{1}x)^{t}+\sum_{k=2}^{N}\left[\sum_{t=1}^{\infty}(\lambda_{k}x)^{t}\right]\psi_{kj}^{2}\notag\\
&=&\pi_{j}\frac{1}{1-x}+\sum_{k=2}^{N}\frac{1}{1-\lambda_{k}x}\psi_{kj}^{2}\,.
\end{eqnarray}
Substituting Eqs.~(\ref{MFPT07}) and~(\ref{MFPT08}) into Eq.~(\ref{MFPT05}) and carrying out some computation, we obtain
\begin{equation}\label{MFPT09}
\tilde{Q}_{ij}(x)=\frac{\pi_{j}+(1-x)\displaystyle\sum_{k=2}^{N}\frac{1}{1-\lambda_{k}x}\psi_{ki}\psi_{kj}\sqrt{\frac{s_{j}}{s_{i}}}}{\pi_{j}+(1-x)\displaystyle \sum_{k=2}^{N}\frac{1}{1-\lambda_{k}x}\psi_{kj}^{2}}\,.
\end{equation}
Differentiating $\tilde{Q}_{ij}(x)$ with respect to $x$ and then setting $x=1$, we obtain a closed-form expression for $F_{ij}$ as
\begin{equation}\label{MFPT11}
F_{ij}=\frac{s}{s_{j}}\sum_{k=2}^{N}\frac{1}{1-\lambda_{k}}\left(\psi_{kj}^{2}-\psi_{ki}\psi_{kj}\sqrt{\frac{s_{j}}{s_{i}}}\right)\,,
\end{equation}
which is 
useful for the following derivation. Similarly to the stationary distribution, the MFPT $F_{ij}$ is also related to the strength of the target node and weights of the edges involved.

\section{MFPT to a given target}

After obtaining the MFPT $F_{ij}$ from one node $i$ to another one $j$ in a generic weighted network $G$, we proceed to consider the trapping problem defined on $G$, which is a particular random walk with a perfect trap placed at a given node, e.g., node $j$. We use $\langle F_{j}\rangle$ to represent the average trapping time, which is the mean of the MFPT $F_{i j}$ from a starting point $i$ to the trap node $j$, with the starting point being taken over the stationary distribution~\cite{TeBeVo09} that depends on the strength, as can be seen from Eq.~(\ref{Trans11}). Note that the selection of the starting point is slightly different from that in most existing works~\cite{KaBa02,Ag08,HaRo08,KiCaHaAr08,ZhQiZhXiGu09,ZhXiZhLiGu09,ZhGuXiQiZh09,AgBu09}, where the distribution of the starting points is uniform. The choice of the stationary distribution for the starting points adopted here is reasonable, since we focus on the trapping problem on weighted networks.

By definition, the ATT $\langle F_{j} \rangle$ is given by~\cite{TeBeVo09}
\begin{equation}\label{ATT01}
\langle F_{j}\rangle=\frac{1}{1-\pi_j}\sum_{i=1}^{N}\pi_i\,F_{ij}\,.
\end{equation}

Taking into account the expression of $F_{ij}$ in Eq.~(\ref{MFPT11}), one straightforwardly arrives at
\begin{eqnarray}\label{ATT02}
\langle F_{j}\rangle &=&\frac{1}{1-\pi_j}\sum_{i=1}^{N}\frac{s_{i}}{s_{j}}\sum_{k=2}^{N}\frac{1}{1-\lambda_{k}}\left(\psi_{kj}^{2}-\psi_{ki}\psi_{kj}\sqrt{\frac{s_{j}}{s_{i}}}\right)\notag\\
&=&\frac{1}{1-\pi_j}\sum_{k=2}^{N}\left(\frac{1}{1-\lambda_{k}}\psi_{kj}^{2}\sum_{i=1}^{N}\frac{s_{i}}{s_{j}}\right)\notag \\&\quad&-\frac{1}{1-\pi_j} \sum_{k=2}^{N}\left(\frac{1}{1-\lambda_{k}}\psi_{kj}\sqrt{\frac{s}{s_{j}}}\sum_{i=1}^{N}\psi_{ki}\sqrt{\frac{s_{i}}{s}}\right).
\end{eqnarray}
Using Eq.~(\ref{Trans03}) gives $\sum_{i=1}^{N}\psi_{ki}\sqrt{\frac{s_{i}}{s}}=\sum_{i=1}^{N}\psi_{ki} \psi_{1i}$=0. Thus, Eq.~(\ref{ATT02}) reduces to
\begin{equation}\label{ATT03}
\langle F_{j}\rangle =\frac{1}{1-\pi_j}\frac{s}{s_j}\sum_{k=2}^{N}\frac{1}{1-\lambda_{k}}\psi_{kj}^{2}\,.
\end{equation}
Equation~(\ref{ATT03}) is a general result valid for the trapping problem on all weighted networks.

We next derive a lower bound for $\langle F_{j}\rangle$. By Cauthy's inequality, one has
\begin{equation}\label{ATT04}
\left(\sum_{k=2}^{N}\frac{1}{1-\lambda_{k}}\psi_{kj}^{2}\right)\left(\sum_{k=2}^{N}(1-\lambda_{k})\psi_{kj}^{2}\right)\geq \left(\sum_{k=2}^{N}\psi_{kj}^{2}\right)^2\,.
\end{equation}
According to Eq.~(\ref{Trans06}), every diagonal entry entry $s_{jj}$ of matrix $S$ satisfies $s_{jj}=\sum_{k=1}^{N}\lambda_k  \psi_{kj}^2\geq 0$ , $j=1,2,\ldots,N$. Considering that the traces of matrices $S$ and $P$ are equivalent to each other, both equaling zero, we have $s_{jj}=0$. Then, applying $\lambda_{1}=1$, one gets
\begin{equation}\label{ATT05}
\sum_{k=2}^{N}(1-\lambda_{k})\psi_{kj}^{2}=\sum_{k=1}^{N}(1-\lambda_{k})\psi_{kj}^{2}=1-\sum_{k=1}^{N}\lambda_{k}\psi_{kj}^{2}=1\,
\end{equation}
and
\begin{equation}\label{ATT06}
\sum_{k=2}^{N}\psi_{kj}^{2}=\sum_{k=1}^{N}\psi_{kj}^{2}-\pi_{j}=1-\pi_{j}\,.
\end{equation}
Combining Eqs.~(\ref{ATT03})-(\ref{ATT06}) produces the following  result:
\begin{equation}\label{ATT08}
\langle F_{j}\rangle \geq \frac{1}{1-\pi_j}\frac{s}{s_j}(1-\pi_j)^2=\frac{s}{s_j}(1-\pi_j)=\frac{s}{s_j}-1.
\end{equation}
This lower bound is strongly affected by the strength of the trapping node and the weights of all edges in the network. In addition, as will be shown in the following section, the lower bound for ATT can be achieved in some graphs.

\section{Results for uncorrelated weighted networks}

We have shown that the primary quantities related to random walks on weighted networks rely on the node strengths and the weights of edges. Below, we will apply the above-obtained results to uncorrelated networks in order to uncover how these primary quantities change with the weight parameter $\theta$.

For uncorrelated networks, many quantities of interest can be determined explicitly. First, for the strength $s_i$ of node $i$, by definition we have
\begin{equation}\label{Uncorr01}
s_{i}=\sum_{j\in \Omega_i}(d_{i} d_{j})^\theta=(d_{i})^\theta\sum_{d'=d_{\rm min}}^{d_{\rm max}}d_{i}P(d'|d_i)( d')^\theta\,,
\end{equation}
where $P(d'|d_i)$ is the conditional probability~\cite{PaVaVe01} that a node of degree $d_i$ has a neighboring node with degree $d'$; $d_{\rm min}$ and $d_{\rm max}$ are the minimum and maximum node degrees, respectively. Let $P(d)$ be the degree distribution of the network. Then, for an uncorrelated network, $P(d'|d_i)=d'P(d')/\langle d \rangle$ and
\begin{equation}\label{Uncorr02}
s_{i}=(d_{i})^{\theta+1}\sum_{d'=d_{\rm min}}^{d_{\rm max}}\frac{(d')^{\theta+1}P(d')}{\langle d \rangle}=\frac{(d_{i})^{\theta+1}\langle d^{\theta+1} \rangle}{\langle d \rangle}\,,
\end{equation}
where $\langle d^{\theta+1} \rangle$ is the $(\theta+1)$th order moment of the degree distribution.

The total strength $s$ of all nodes can be evaluated as
\begin{equation}\label{Uncorr03}
s=NP(s_i)s_{i}=NP(d_i)s_{i}=\frac{N \langle d^{\theta+1} \rangle^2}{\langle d \rangle}\,,
\end{equation}
where $P(s_i)$ is the distribution of node strengths that is equal to the degree distribution of the network. Thus, the stationary distribution for a random walk on an uncorrelated weighted network becomes
\begin{eqnarray}\label{Uncorr04}
\pi&=&\left(\frac{s_{1}}{s}, \frac{s_{2}}{s},\ldots, \frac{s_{N}}{s}\right)^\top \nonumber \\
&=&\left(\frac{(d_{1})^{\theta+1}}{N \langle d^{\theta+1} \rangle}, \frac{(d_{2})^{\theta+1}}{N \langle d^{\theta+1} \rangle},\ldots, \frac{(d_{N})^{\theta+1}}{N \langle d^{\theta+1} \rangle}\right)^\top \,.
\end{eqnarray}

Equation~(\ref{Uncorr04}) shows that the stationary distribution is dominated by the weight parameter $\theta$. When $\theta=0$, it reduces to the case of unbiased random walks. For the case of $\theta=-1$, the stationary distribution is uniform with $\pi_i=1/N$ for node $i$. In the case of $\theta>-1$, $\pi_i$ increases with $d_{i}$, which means that it is easier to find the walker after a long time at a large-degree node than at a small-degree node. On the contrary, in the case of $\theta<-1$, $\pi_i$ is a decreasing function of $d_{i}$, and thus the final occupation probability for a node with a small degree is higher than another node with a large degree.

From Eq.~(\ref{Uncorr04}), we can also obtain the mean first return time (MFRT) $R_{ii}$ for uncorrelated weighted networks. By definition, MFRT $R_{ii}$ is the expected time for a walker originating from node $i$ and returning to $i$ for the first time. According
to the Kac formula~\cite{AlFi99,CoBeMo07,SaDoMe08}, the MFRT for a node
coincides with the inverse probability to find the walker at
this node in the final equilibrium state of the random-walk process. Thus,
\begin{equation}\label{Uncorr05}
R_{ii}=\frac{1}{\pi_i}= \frac{N \langle d^{\theta+1}\rangle}{(d_{i})^{\theta+1}} \,,
\end{equation}
which also depends on $\theta$ and has been derived in~\cite{FrFR09} by using a different approach.

Finally, for the lower bound of ATT on an uncorrelated network, inserting Eqs.~(\ref{Uncorr02}) and~(\ref{Uncorr03}) into Eq.~(\ref{ATT08}) yields
\begin{equation}\label{Uncorr06}
\langle F_{j}\rangle \geq \frac{N \langle d^{\theta+1}\rangle}{(d_{j})^{\theta+1}}-1.
\end{equation}
This lower bound is sharp. It can be reached in an $N$-node complete graph, where the MFPT from any node to $j$ equals $N-1$~\cite{Bobe05}, which is consistent with Eq.~(\ref{Uncorr06}).
Moreover, the ATT for complete graphs is independent of parameter $\theta$.

Different from complete graphs and some other regular graphs belonging to circular graphs, the lower bound for ATT provided by Eq.~(\ref{ATT08}) on non-regular networks relies heavily on the parameter $\theta$. For those networks with small relaxation time when the stationary distribution is reached, the lower bound can be attained (see~\cite{LaSz10} for explanation) but the results for different destinations are closely related to $\theta$. For $\theta>-1$, the ATT for large-degree nodes is small, since $F_{j}$ decreases with growing $d_j$. For $\theta=-1$, the ATT is identical for all destination nodes irrespective of their degrees, while for $\theta<-1$, the walker can find small-degree nodes more easily than finding large-degree nodes, since in this case the walker is biased toward those nodes with small degrees.

Note that for the $\theta=0$ case, it is reduced to unbiased random walks. For this particular case, Eq.~(\ref{ATT08}) shows that the lower bound of ATT is proportional to the inverse degree of the trap node. Such a scaling has been reported in~\cite{TeBeVo09} and~\cite{LaSz10}. In~\cite{TeBeVo09}, a main concern is ATT on scale-free networks, which are ubiquitous in real systems~\cite{AlBa02,Ne03}.

Next, we show that our results for ATT given in Eq.~(\ref{ATT08}) can display various different scalings when the trap is located on a target node having a maximum degree on a scale-free network and provide important information about random walks on the scale-free network, encompassing that in~\cite{TeBeVo09} as a particular case.

For an uncorrelated scale-free network with a constant average degree $\langle d \rangle$ and a power-law degree distribution $P(d)\sim d^{-\gamma}$, 
the term $\langle d^{\theta+1}\rangle$ in Eq.~(\ref{Uncorr06}) can be evaluated as follows: for $\theta=-1$, $\langle d^{\theta+1}\rangle=1$; for $\theta=0$, $\langle d^{\theta+1}\rangle=\langle d \rangle$ is exactly the average node degree; while for other cases ($\theta \neq 0\,,-1$), by integrating $d^{\theta+1} P(d)$ from $d_{\rm min}$ to $d_{\rm max}$ with respect to $d$, we obtain $\langle d^{\theta+1}\rangle$ to be approximatively
\begin{equation}\label{Uncorr07}
\langle d^{\theta+1}\rangle \sim  \begin{cases}
\ln d_{\rm max}, &\theta=\gamma-2, \\
(d_{\rm max})^{\theta-\gamma+2}, & \theta > \gamma-2,\\
(d_{\rm min})^{\theta-\gamma+2}, & \theta < \gamma-2.
\end{cases}
\end{equation}

On the other hand, it has been established~\cite{DoGoMe08} that for an uncorrelated scale-free network with $\gamma \geq 3$, its
$d_{\rm max}$ approximately obeys the relation $d_{\rm max}\sim N^{1/(\gamma-1)}$. Thus, when a trap is fixed at a hub node having degree $d_{\rm max}$ on such a network with $\gamma \geq 3$, the lower bound of ATT, $\langle F_{d_{\rm max}}\rangle$, can be expressed in terms of the network size $N$. For $\theta=-1$, $\langle F_{d_{\rm max}}\rangle=N-1$, which scales linearly with $N$. In fact, for this special case, the ATT is independent of the network type and the trap position. For $\theta=0$, $\langle F_{d_{\rm max}}\rangle=N^{(\gamma-2)/(\gamma-1)}$, growing sublinearly with $N$ as observed in~\cite{TeBeVo09}. For $\theta \neq 0\,,-1$, we distinguish three cases:
\begin{equation}\label{Uncorr08}
\langle F_{d_{\rm max}}\rangle \sim \begin{cases}
\ln N, &\theta=\gamma-2, \\
{\rm 1}, & \theta > \gamma-2,\\
N^{1-\frac{(\gamma-2)(\theta+1)}{(\gamma-1)}}, & \theta < \gamma-2\,.
\end{cases}
\end{equation}
Note that the approximate formulas given in Eq.~(\ref{Uncorr08}) are valid for small $\theta$. Particularly, when $\theta=0$, Eq.~(\ref{Uncorr08}) is in agreement with the previous result obtained in~\cite{TeBeVo09}.  For other $\theta$, Eq.~(\ref{Uncorr08}) may yield a large deviation from the true values~\cite{BaPa10}.

\begin{figure}
\begin{center}
\includegraphics[width=1.1\linewidth,trim=0 50 0 50]{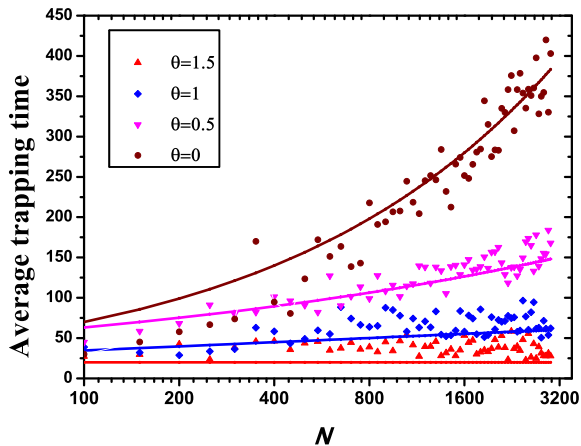}
\end{center}
\caption[kurzform]{\label{trapping} (Color online) Average trapping time to a hub node for biased random walks on the Barab\'asi-Albert scale-free network for various $\theta$ and network size $N$. The symbols represent the numerical results generated by Eq.~(\ref{ATT02}), while the solid lines stand for the corresponding analytical results given by Eq.~(\ref{Uncorr08}).}
\end{figure}

To confirm our analytical results given in Eq.~(\ref{Uncorr08}), we perform extensive simulations of biased random walks on the Barab\'asi-Albert (BA) scale-free network~\cite{BaAl99}, the degree distribution exponent $\gamma$ of which is equal to 3. In Fig.~\ref{trapping}, we present our numerical results for the ATT to a most-connected node of BA networks with average node degree 4, which confirm the theoretical prediction provided by  Eq.~(\ref{Uncorr08}).

Equation~(\ref{Uncorr08}) shows that, for large $\theta$, one can find the trap hub easily. For example, in the whole range of $-1 < \theta \leq 0$, when $\theta$ is not very large, the ATT increases sublinearly with $N$, displaying a high trapping efficiency. Actually, for some particular $\theta$ this result is also valid for correlated scale-free networks. Previous studies have indicated that, in the case of $\theta = 0$, the ATT to a hub node in some correlated scale-free networks, such as the pseudofractal web~\cite{ZhQiZhXiGu09,ZhXiZhLiGu09}, the Apollonian network~\cite{ZhGuXiQiZh09}, hierarchical scale-free networks~\cite{AgBu09,AgBuMa10,MeAgBeVo12}, and modular scale-free networks~\cite{ZhLiGaZhGuLi09,ZhYaLi12}, behaves sublinearly with $N$.
For small $\theta$, especially for $\theta < -1$, Eq.~(\ref{Uncorr08}) shows that the walker will miss the hub node when $\theta$ decreases to a sufficiently small value, since in this case the walker always travels among some nodes with small degrees.

\section{Conclusions}

In this paper, we have presented a unified framework for random walks on weighted networks. In the process of random walks, the transition probability from node $i$ to node $j$ is proportional to the weight of the edge connecting $i$ and $j$, which is  $(d_{i}d_{j})^\theta$, where $d_{i}$ and $d_{j}$ are, respectively, the degrees of $i$ and $j$, and $\theta$ is a control parameter of weights. We have calculated analytically the expressions of the stationary distribution, MFPT from an arbitrary node to any other node, and ATT to a given trap node with a lower bound obtained. The resultant formulas for all these interesting quantities are sensitive to the change of $\theta$, implying that $\theta$ plays an essential role for random walks on weighted networks. For the special case of uncorrelated networks, we have derived a succinct expression for the stationary distribution, based on which we have further shown how the lower bound of ATT scales with the network size. Particularly, we have provided a comprehensive analysis for ATT on uncorrelated scale-free networks, when a trap is positioned at a hub node. Our work provides an efficient method for controlling random-walk dynamics; using this method one can accelerate or decelerate the diffusion process on a weighted network.

\begin{acknowledgments}
We thank Pengcheng Li for his assistance. This work was supported by the National Natural Science Foundation of China under Grants No. 61074119 and No. 11275049 and the Hong Kong Research Grants
Council under the GRF Grant CityU 1114/11E.
\end{acknowledgments}

\nocite{*}


\end{document}